\documentclass[aps,prl,superscriptaddress,twocolumn]{revtex4-2}
\usepackage{graphicx} 
\usepackage{subfigure}
\usepackage{multirow}
\usepackage{xspace}
\usepackage{fancyhdr}
\usepackage{amsmath,amssymb}
\usepackage{xcolor}

\begin{document}

\title{Revealing the nature of the charge density wave order of ErTe$_3$ via Raman scattering under anisotropic strain}

\author{Théotime Freitas}
\email{theotime.freitas@u-paris.fr}
\affiliation{Université Paris Cité, CNRS, Laboratoire Matériaux et Phénomènes Quantiques, Paris, France}
\author{Mattia Udina}
\email{mattia.udina@gmail.com}
\affiliation{Université Paris Cité, CNRS, Laboratoire Matériaux et Phénomènes Quantiques, Paris, France}
\affiliation{Institut de Physique et Chimie des Mat\'{e}riaux de Strasbourg (UMR 7504),  Universit\'{e} de Strasbourg and CNRS, Strasbourg, 67200, France}
\author{Alexandr Alekhin}
\affiliation{Université Paris Cité, CNRS, Laboratoire Matériaux et Phénomènes Quantiques, Paris, France}
\author{Niloufar Nilforoushan}
\affiliation{Université Paris Cité, CNRS, Laboratoire Matériaux et Phénomènes Quantiques, Paris, France}
\author{Sarah Houver}
\affiliation{Université Paris Cité, CNRS, Laboratoire Matériaux et Phénomènes Quantiques, Paris, France}
\author{Alain Sacuto}
\affiliation{Université Paris Cité, CNRS, Laboratoire Matériaux et Phénomènes Quantiques, Paris, France}
\author{Benito A. Gonzalez}
\affiliation{Department of Applied Physics, Stanford University, Stanford, CA, USA}
\affiliation{Geballe Laboratory for Advanced Materials, Stanford University, Stanford, California 94305, USA}
\affiliation{Stanford Institute for Materials and Energy Sciences, SLAC, Menlo Park, California 94025, USA}
\author{Ian R. Fisher}
\affiliation{Department of Applied Physics, Stanford University, Stanford, CA, USA}
\affiliation{Geballe Laboratory for Advanced Materials, Stanford University, Stanford, California 94305, USA}
\affiliation{Stanford Institute for Materials and Energy Sciences, SLAC, Menlo Park, California 94025, USA}
\author{Indranil Paul}
\email{indranil.paul@u-paris.fr}
\affiliation{Université Paris Cité, CNRS, Laboratoire Matériaux et Phénomènes Quantiques, Paris, France}
\author{Yann Gallais}
\email{yann.gallais@u-paris.fr}
\affiliation{Université Paris Cité, CNRS, Laboratoire Matériaux et Phénomènes Quantiques, Paris, France}
\affiliation{Institut Universitaire de France (IUF)}

\begin{abstract}
The nature of the charge density wave (CDW) order of the rare-earth tritelluride ErTe$_3$ is investigated by Raman scattering under anisotropic strain. The CDW state of ErTe$_3$ is unconventional since
it is accompanied by an unusual mirror symmetry breaking, whose origin remains to be understood. Studying the polarization-resolved Raman spectrum of the CDW amplitude mode as a function of strain and temperature, we find that the mirror symmetry breakings of the CDW state are not independent, arguing against the recently proposed ferro-axial multi-component CDW order. Instead, we show that a single component CDW order parameter with an ordering wavevector tilted away from the principle crystallographic axis can reproduce the observed mirror symmetry breakings and their manifestation in the Raman spectra.
\end{abstract}

\maketitle



A central step to describe any symmetry-broken state is to identify its order parameter. This task can be non-trivial when the ordered state breaks additional symmetries beyond those expected on general grounds. An example of this complexity are the rare-earth tritellurides (RTe$_3$) which have long been regarded as prototypical charge density wave (CDW) materials \cite{dimasi_chemical_1995,ru_effect_2008,yao_theory_2006,brouet_angle-resolved_2008,johannes_fermi_2008,singh_2025}. Yet, recent experiments suggest that the CDW state in RTe$_3$ breaks additional mirror symmetries, pointing to a richer form of CDW order than previously assumed \cite{wang_axial_2021,singh_ferroaxial_2025,wulferding_magnetic_2025}. Determining the nature of the order parameter is an important conceptual challenge, with implications extending beyond RTe$_3$ to other CDW systems.

In general, while the CDW order parameter $\phi$ itself can be 
without internal structure and thus preserve all mirror operations, the mirror symmetry can be broken by the CDW wavevector ${\bf Q}_{CDW}$ itself. However, even when $\phi$ has internal structure the CDW order is not expected to break the mirror symmetries 
which map ${\bf Q}_{CDW} \rightarrow -{\bf Q}_{CDW}$ [note, $\phi(-{\bf Q}_{CDW}) = 
\phi^{*}({\bf Q}_{CDW})$, and the physical variables depend on $|\phi({\bf Q}_{CDW})|$].

\begin{figure*}[t]
    \centering
    \includegraphics[width=18cm]{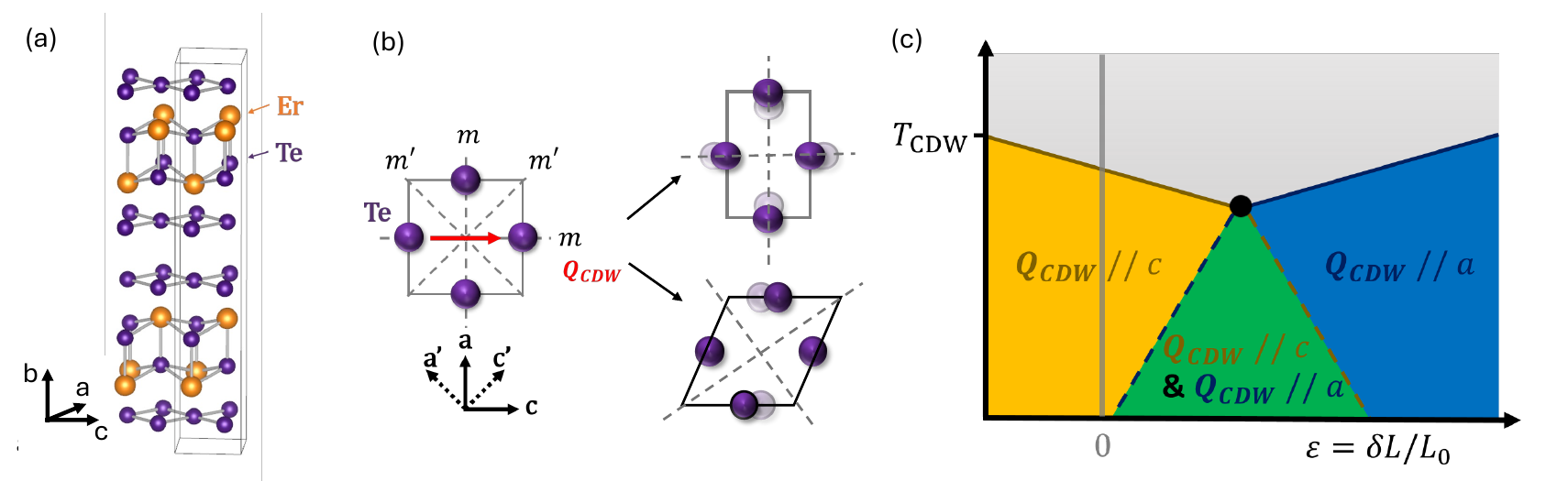}
    \caption{(a) Crystal structure of ErTe$_3$ and (b) mirror symmetries of the Te atom square $ac$ plane. Note that for orthorhombic point groups, the long axis is along the $b$ direction. The CDW ordering wave-vector ${\bf Q}_{CDW}$ of pristine ErTe$_3$ is along the $c$-axis and thus breaks $m^{\prime}$ mirrors. The sketch of the Te plane distortions illustrates $m$ (below) and $m^{\prime}$ (above) symmetry breakings of the Te atom square plane. (c) Expected temperature-strain phase diagram of ErTe$_3$, for a strain along the $a$ axis, based on previous transport and X-ray measurements. It shows a bi-critical point where the CDW ordering transitions along orthogonal directions $c$ and $a$ meet \cite{straquadine_evidence_2022,gallofrantz_charge_2024,singh_emergent_2024}. When varying strain at fixed temperature, the transition between the two CDW orderings occurs via an intermediate region (shaded in green) where both orders co-exist. The pristine $\epsilon=0$ ErTe$_3$ is marked by a vertical grey line. Note that the second phase transition to a homogeneous bi-directional CDW state found at lower temperature in ErTe$_3$ is not displayed \cite{ru_effect_2008}. It is distinct from the green shaded phase depicted here.}
        \label{fig1}
\end{figure*}

In nearly tetragonal RTe$_3$, ${\bf Q}_{CDW}$ is thought to be along the cristallographic axis $c$, which evidently breaks the $m^{\prime}$ mirrors of the idealized square Te net but preserves the $m$ mirror along $a$ (Fig.\ \ref{fig1}(a) and (b)). Recent Raman and second harmonic generation measurements however \cite{wang_axial_2021,singh_ferroaxial_2025,wulferding_magnetic_2025}, indicate a CDW order that breaks all mirror reflections of the Te square plane, consistent with a monoclinic rather than an orthorhombic point group symmetry. These findings have put into question the precise nature of the CDW order parameter of RTe$_3$.
\par
Here we address this conundrum using elasto-Raman scattering on ErTe$_3$ under an applied uniaxial stress $\sigma_{aa}$ along the $a$ axis. Symmetry-resolved Raman response is a particularly powerful tool to study this problem since it allows keeping track of two quantities $\Sigma_{m^{\prime}}$ and $\Sigma_{m}$ (defined later) which monitor $m^{\prime}$- and $m$-mirror symmetry breakings, respectively. Tracking both quantities, as functions of temperature and applied stress, allows us to determine whether
these two symmetry breakings are independent or intrinsically linked. Furthermore, uniaxial stress along $a$ is a very useful tuning parameter for two reasons. First, it couples directly to $\Sigma_{m^{\prime}}$ but not to $\Sigma_{m}$, thus, clearly distinguishing the two. Second, if $m$ and $m^{\prime}$ symmetry breaking are distinct component of a multi-component CDW order parameter as proposed recently~\cite{alekseev_charge_2024}, then we expect to see two transitions under strain \cite{SI}. 
Instead, we show that both symmetry breakings track each other as a function of both strain and temperature strongly supporting a single-component order parameter $\phi({\bf Q}_{CDW})$. We further propose a scenario where the mirror symmetry breakings are associated with a ${\bf Q}_{CDW}$ that is tilted away from the principal crystallographic axes. Calculations of the symmetry-resolved Raman response within this scenario are found to reproduce the main features of the experimental data.

 \begin{figure}[b]
    \centering
    \includegraphics[width=8cm]{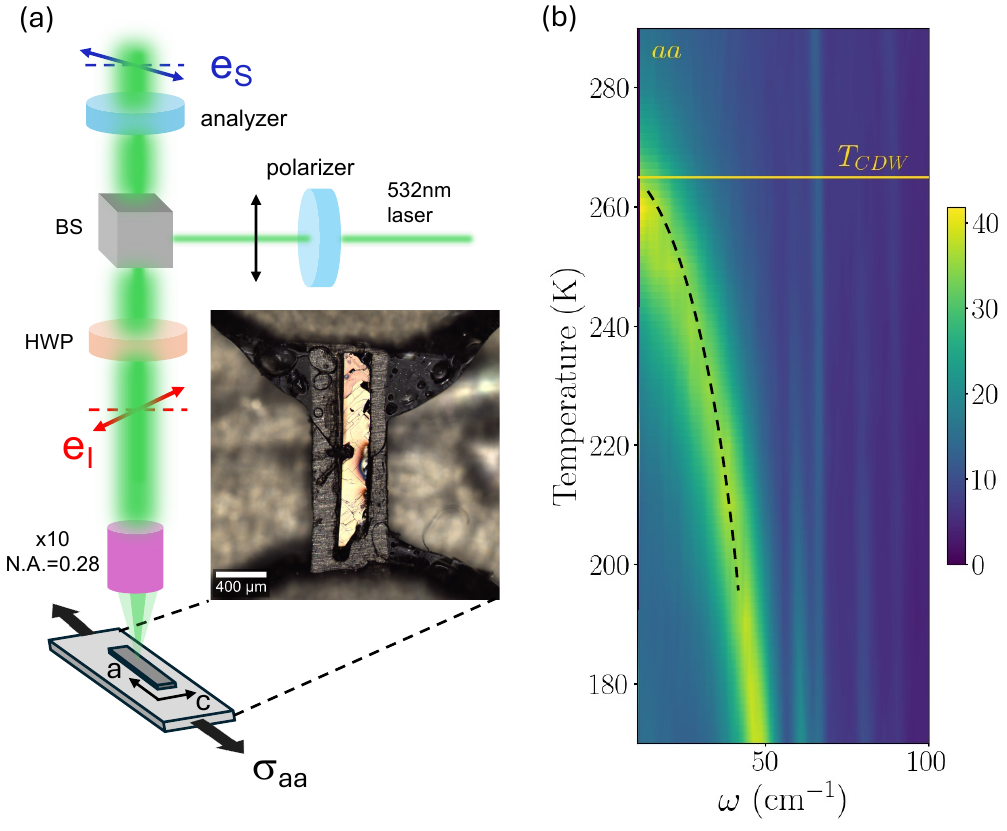}
    \caption{(a) Polarization-resolved Raman scattering set-up under uni-axial stress. $e_i$ and $e_s$ indicate the polarization of the incident and collected light, respectively. BS: beam-splitter, HWP: half-wave plate. A thin ErTe$_3$ crystal is glued on a titanium plate on which the uni-axial stress, $\sigma_{aa}$, is applied along its $a$ axis. The inset shows an optical image of the crystal mounted on the Ti strain platform. (b) Color plot of the temperature dependent Raman spectrum of an unstrained crystal of ErTe$_3$ across the CDW phase transition ($T_{CDW}\sim$ 265~K) in $aa$ polarization configuration where both incoming and outgoing photon polarizations are along the $a$ axis. The temperature dependence of the CDW amplitude mode is marked by a dashed line which is a guide to the eye.}
    \label{fig2}
\end{figure}


The RTe$_3$ compounds consist of a stacking of square Te planes ($ac$ planes) with a glide-plane that weakly breaks the $m^{\prime}$ mirror symmetry (see Fig.\ \ref{fig1}(a) and (b)). The presence of the glide plane breaks the in-equivalence between the CDW ordering along $c$ and $a$ axis, and favours one direction for the CDW ordering at the expense of the other \cite{yao_theory_2006}. The nearly degenerate CDW orderings make these materials an ideal playground for anisotropic strain control of electronic orders \cite{hicks_probing_2025}. This was demonstrated by recent transport and X-ray measurements where modest applied strains could switch the direction of the CDW ordering \cite{straquadine_evidence_2022,gallofrantz_charge_2024,singh_emergent_2024}. The qualitative temperature-strain phase diagram is depicted in Fig.\ \ref{fig1}(c) \cite{straquadine_evidence_2022,gallofrantz_charge_2024,singh_emergent_2024}.

Variable temperature elasto-Raman scattering experiments were performed on a single crystal of ErTe$_3$ which undergoes a CDW phase transition at $T_{CDW}\sim265$~K. In situ mechanical strain was applied to the sample using a piezo-based strain cell. The ErTe$_3$ crystal was glued on a titanium (Ti) platform on which both compressive and tensile stresses were applied (Fig.\ \ref{fig2}(a)). The crystal was oriented so that the stress was applied along the $a$ axis. In this configuration the sample experiences in-plane strains $\epsilon_{xx}$ and $\epsilon_{yy}$. A slight misalignment of the crystal axis with respect to the Ti platform meant that a small residual shear strain $\epsilon_{xy}$, estimated to be less than 5 percents of $(\epsilon_{xx} - \epsilon_{yy})$, was also applied to the sample (here and throughout the paper we take the $x$ and $y$ axis along the crystallographic $a$ and $c$ axis respectively). Polarization-resolved Raman measurements were performed using a 532~nm solid state laser in back-scattering geometry with co-linear incoming and outgoing photon wave-vectors. The laser beam was focused using a x10 long working distance objective and the spot diameter was about 6~$\mu m$. The optical set-up is displayed in Fig.\ \ref{fig2}(a). The deformation of the Ti platform along the applied stress was defined as $\varepsilon=\frac{\delta L}{L_0}$ where $L_0$ is the initial length of the Ti platform and $\delta L$ its variation. It was monitored via a capacitance sensor affixed to the moving plates. In general we do not expect full transmission of the strain to the sample and the quoted nominal strains $\varepsilon$ are over-estimation of the actual strain (see SI for more details about the strain device and the Raman set-up \cite{SI}).

\par
\begin{figure*}[t]
    \centering
    \includegraphics[width=18cm]{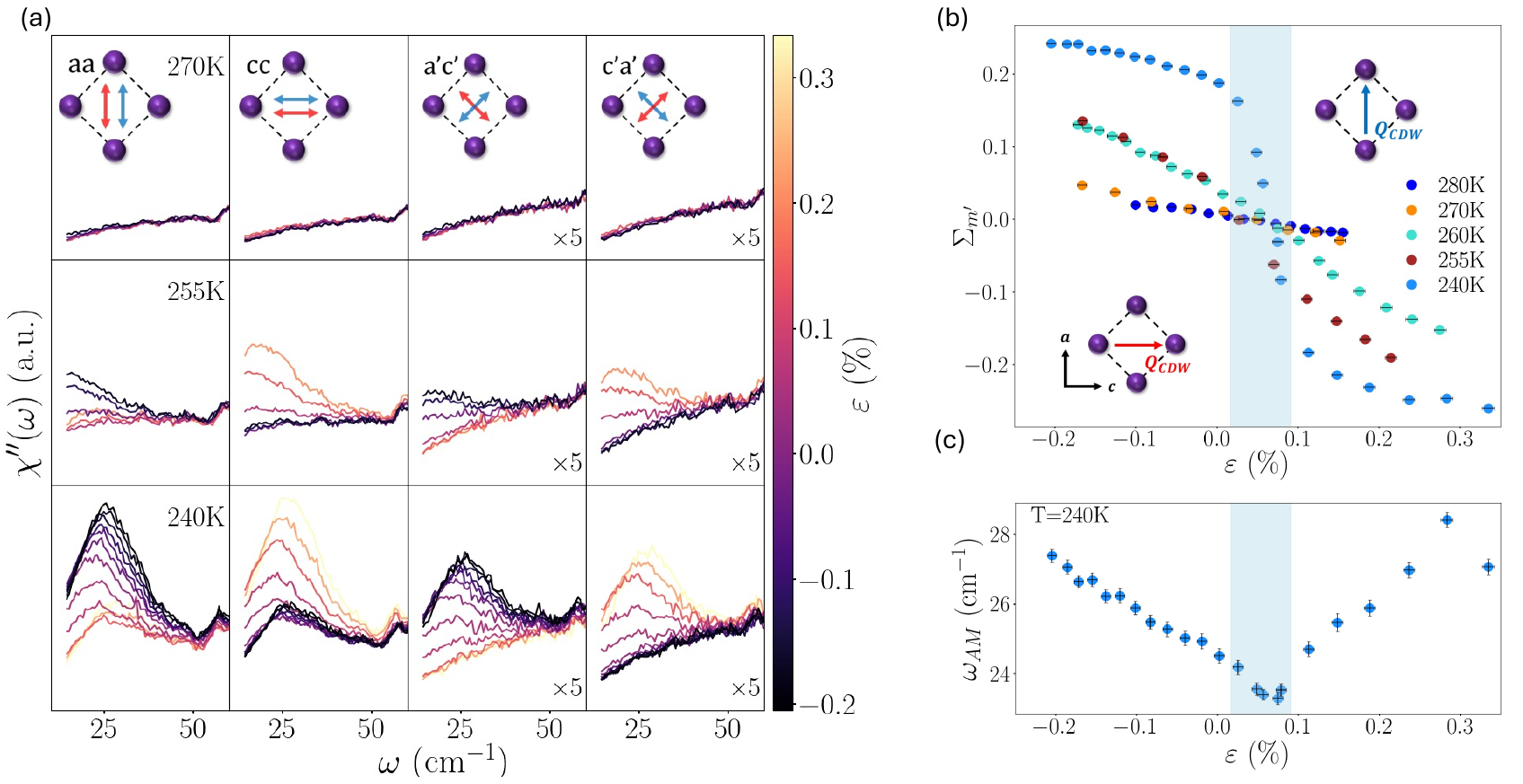}
    \caption{(a) Strain dependence of the low energy Raman response $\chi''$ for several temperatures and different polarization configurations: $aa$, $cc$, $a'c'$ and $c'a'$ (see insets). The Raman response $\chi''$ was obtained from the raw Raman intensity by correcting it with the Bose factor \cite{SI}. The color bar indicates the nominal strain values along the $a$ axis of the crystal: negative for compressive strain, positive for tensile strain. (b) Strain dependence of $\Sigma_{m^{\prime}}$ for 5 different temperatures. The sign switch of $\Sigma_{m^{\prime}}$ is associated with the rotation of $Q_{CDW}$ which is aligned along $c$ under compression and aligned along $a$ under tension. The blue shaded area for weak tensile strain indicates the regime where domains with different orientation of $Q_{CDW}$ co-exist. (c) Evolution of the CDW AM energy as a function of strain at T=240~K.}
    \label{fig3}
\end{figure*}
Figure \ref{fig2}(b) shows a color plot of the temperature dependent Raman spectrum of ErTe$_3$ across $T_{CDW}$ under zero applied strain, $\epsilon$=0, using the polarization configuration $aa$ (incoming and outgoing polarization along the $a$ axis). As already reported previously the CDW state is characterized by the emergence of the amplitude mode (AM) of the CDW order below $T_{CDW}$ \cite{lavagnini_raman_2010,eiter_alternative_2013,lazarevic_evidence_2011} (dashed line in Fig. \ref{fig2}(b)). Its energy, close to $\sim$ 50~cm$^{-1}$ at 170~K, softens upon approaching $T_{CDW}$, consistent with a second order phase transition. In the following, we concentrate on the polarization dependence of the CDW AM, which is a sensitive probe of the symmetry of the underlying CDW order.

 \begin{figure*}[t]
    \centering
    \includegraphics[width=17cm]{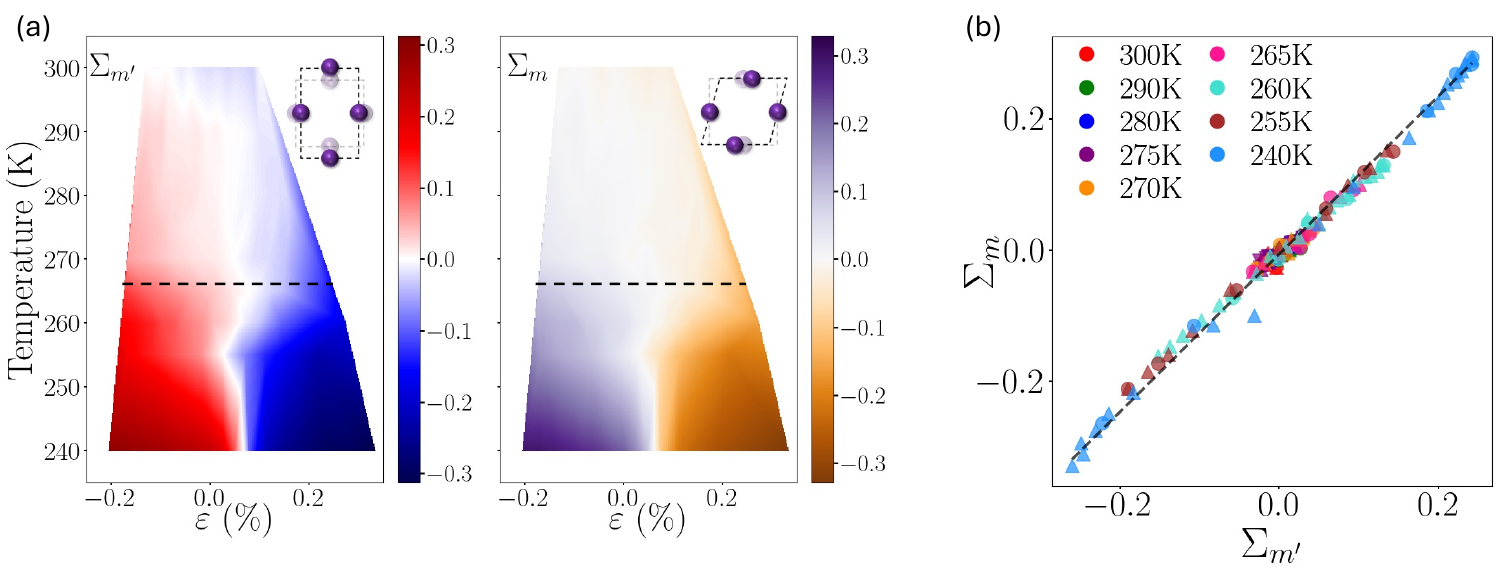}
    \caption{(a) Color plot of the temperature strain phase diagram of $\Sigma_{m^{\prime}}$ and $\Sigma_m$. It was obtained by extrapolating measurements at 8 different temperatures with at least 10 strain values for each temperature. $T_{CDW}\sim $265~K is marked by a dashed line. (b) Linear scaling of $\Sigma_{m^{\prime}}$ and $\Sigma_m$ using strain and temperatures as implicit parameters. }
    \label{fig4}
\end{figure*}
\par
Figure \ref{fig3}(a) shows the polarization-resolved Raman response $\chi''$ of ErTe$_3$ as a function of the applied strain $\varepsilon$ along the $a$ axis for three selected temperatures above and below $T_{CDW}$. The four different polarizations configurations are depicted in the insets. For T=270~K, above the pristine $T_{CDW}\sim$ 265~K, the spectra display a featureless continuum typical of a metallic state and do not display any significant strain dependence in all polarization configurations. Below $T_{CDW}$, for both T=255~K and T=240~K, the AM emerges in all configurations and its overall intensity and lineshape display a profound strain dependence. Focusing first on the spectra in $aa$ and $cc$ configurations, we notice that the strain dependence of the AM intensity evolves in opposite ways in the two configurations. Noting $I^{}_{aa}$ and $I^{}_{cc}$ the intensities of the AM in the $aa$ and $cc$ configurations we have $I^{}_{cc} <  I^{}_{aa}$ for compressive strain $\varepsilon<0$ and $I^{}_{cc} > I^{}_{aa}$ for tensile strain $\varepsilon>0$. The strain-induced switch between $I^{}_{cc}$ and $I^{}_{aa}$ is interpreted as a rotation of ${\bf Q}_{CDW}$, with ${\bf Q}_{CDW}\parallel c$ for strong compressive strain and ${\bf Q}_{CDW} \parallel a$ for strong tensile strain. Interestingly a similar switching is also observed between the $a'c'$ and $c'a'$ configuration, a point which will be further discussed below. 
\par

The effect of strain on the orientation of the CDW ordering can be analyzed via the following quantity:
\begin{equation}
\Sigma_{m^{\prime}}=\frac{I^{}_{aa}-I^{}_{cc}}{I^{}_{aa}+I^{}_{cc}}
 \end{equation}
 where $I^{}_{aa/cc}$ are obtained by integrating the Raman intensity of the low energy continuum in the spectral range 0 - 60~cm$^{-1}$ both below and above $T_{CDW}$. Note that $\Sigma_{m^{\prime}}$ changes sign upon the mirror reflection operation $a \to c$ and $c \to a$. It thus quantifies the diagonal mirror $m^{\prime}$ symmetry breaking induced by the orientation of ${\bf Q}_{CDW}$. The evolution of this quantity as a function of strain is shown in Fig.\ \ref{fig3}(b) for several temperatures. Below $T_{CDW}$, the strain induced re-orientation of ${\bf Q}_{CDW}$ manifests itself by a rapid change from $\Sigma_{m^{\prime}}>0$ to $\Sigma_{m^{\prime}}<0$ around $\varepsilon_0\sim 0.1 \%$. Beyond the switching region, $\Sigma_{m^{\prime}}$ saturates for both increasing compressive and tensile strain. The finite value of $\varepsilon_0$ is due to the weak $m^{\prime}$ symmetry breaking associated with the structural glide plane that must be compensated by the applied strain. Spatially-resolved Raman measurements shown in the SI \cite{SI} indicate homogeneous single-domain states across the crystal for saturating strain values. By contrast, a strong spatial dependence of $\Sigma_{m^{\prime}}$ is observed in the intermediate region, close to $\varepsilon_0$. This clearly indicates the presence of domains of different CDW orientations close to the degeneracy point $\varepsilon_0$ where the two phases ${\bf Q}_{CDW} \parallel c$ and ${\bf Q}_{CDW} \parallel a$ meet. This observation validates the proposed phase diagram depicted in Fig.\ \ref{fig1}(c) where the switching between the two ${\bf Q}_{CDW}$ orientations passes through a first order transition and the formation of a spatially inhomogeneous state as in an Ising transition in a magnetic field. In addition, below $T_{CDW}$ the energy of the AM mode displays a pronounced and essentially strain symmetric hardening (Fig.\ \ref{fig3}(c)) indicating a strengthening of the CDW order for both compressive and tensile strains \cite{straquadine_evidence_2022,singh_emergent_2024}.
 \par
  We now study the effect of strain on the more subtle additional $m$ mirror symmetry breaking. For this we define a second symmetry resolved quantity:
 \begin{equation}
  \Sigma^{}_m=\frac{I_{a'c'}^{}-I_{c'a'}^{}}{I_{a'c'}^{}+I_{c'a'}^{}}.
 \end{equation}
This anti-symmetric quantity changes sign upon $m$ mirror reflection and is thus a marker of the $m$ symmetry breaking \cite{singh_ferroaxial_2025,udina_antisymmetric_2025} (see Fig. \ref{fig1}(a)). Similar to  $\Sigma_{m^{\prime}}$, it was obtained by integrating the corresponding Raman spectra between 0 and 60cm$^{-1}$. The parallel evolutions of $\Sigma_{m^{\prime}}$ and $\Sigma_m$ as a function of both strain and temperature are compared in color plots of Fig.\ \ref{fig4}(a). We note that, whereas the sign of $\Sigma_{m^{\prime}}$ is fixed by the orientation of $\phi({\bf Q}_{CDW})$, this is not, a priori, the case for $\Sigma_m$ whose sign for a given $\phi({\bf Q}_{CDW})$ orientation depends only on the orientation of the $m$ symmetry breaking. The color-plots reveal that both quantities display qualitatively similar temperature and strain dependencies. Above $T_{CDW}$ they display rather weak strain dependence. Below $T_{CDW}$ they both switch sign under strain indicating that the re-orientation $\phi({\bf Q}_{CDW})$ is accompanied by a flip of the $m$ symmetry breaking. Spatially-resolved measurements (not shown) do not show any significant variation of $\Sigma_m$ on the $\mu m$ scale for strong tension and compression, indicating a single domain throughout the sample. We attribute this behaviour to the residual $\epsilon_{xy}$ component of the applied strain which favors a specific orientation of the $m$ symmetry breaking (see SI for a discussion of the orientation of monoclinic domains under strain \cite{SI}).

  \begin{figure*}[t]
    \centering
    \includegraphics[width=15cm]{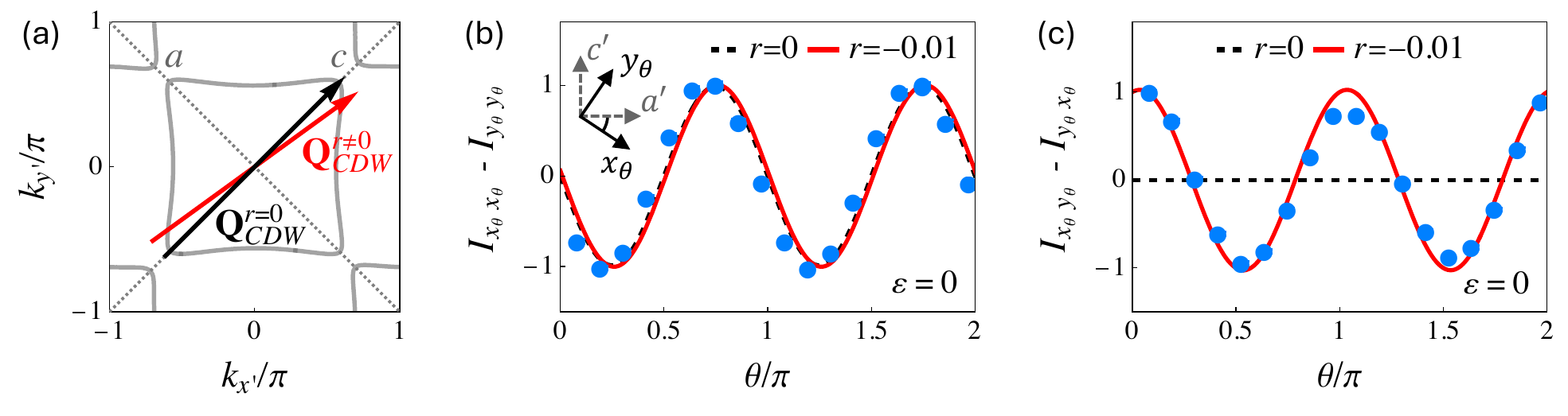}
    \caption{(a) Tight-binding Fermi surface of ErTe$_3$ along with the ${\bf Q}_{CDW}$ either along ($r=0$, black arrow) or tilted from the $c$-axis (red arrow). (b)-(c) Angular dependence of $I_{x_{\theta}x_{\theta}} - I_{y_{\theta}y_{\theta}}$ and $I_{x_{\theta}y_{\theta}} - I_{y_{\theta}x_{\theta}}$, normalized to their respective maximum values, obtained on an un-strained ErTe$_3$ crystal ($\varepsilon = 0$) at 175~K by rotating the incident (scattered) light polarization direction $\hat x_\theta$ ($\hat y_{\theta}$) by an angle $\theta$ relative to the $a'$ ($c'$) axis (inset). Experimental data (blue dots) are compared with theoretical predictions with $r=0$, i.e.\ where ${\bf Q}_{CDW}$ is perfectly aligned along the $c$ axis (dashed lines), and with $r=-0.01$, where ${\bf Q}_{CDW}$ slightly deviates from the high-symmetry direction (solid lines).}
    \label{fig5}
\end{figure*}
 \par

 Further insight into how strain affects both symmetry breakings can be obtained by plotting $\Sigma_{m^{\prime}}$ as a function of $\Sigma_m$ using temperature and strain as implicit parameters as shown in Fig.\ \ref{fig4}(b). Strikingly, $\Sigma_{m^{\prime}}$ and $\Sigma_m$ exhibit almost perfect linear relationship for all applied strains and temperatures. This remarkable finding indicates an intimate link between both mirror symmetry breakings and the $CDW$ order and suggests that they are 
 not independent quantities. This observation, together with the absence of two split phase transitions under substantial strain, implies that the underlying order parameter is a single component and not two-component as suggested by Refs.\ \cite{alekseev_charge_2024,singh_ferroaxial_2025}. In a single component scenario both $m$- and $m^{\prime}$-mirrors are broken at
 $T_{CDW}$, and the same non-zero order parameter contributes to 
$\Sigma_{m^{\prime}}$ and $\Sigma_m$, making them proportional to 
 each other. As further discussed in the SI \cite{SI}, the only way to reconcile a two-component scenario with our data would be to postulate that both temperature \textit{and} strain dependencies of the two components have to be identical (accidental degeneracies), implying microscopic fine tuning. We note that a similar situation has been discussed in the context of the putative multi-component superconducting order parameter of Sr$_2$RuO$_4$ \cite{kivelson_proposal_2020}.
 
\par

The puzzle of simultaneous $m$- and $m^{\prime}$-mirror symmetry breaking can be resolved by recalling that the CDW transition breaks mirrors that are neither parallel nor perpendicular to the ordering wave-vector ${\bf Q}_{CDW}$. In other words, our data suggest that the ordering wave-vector ${\bf Q}_{CDW}$ is not along the high symmetry direction $c$, but is tilted away from it. As shown in the SI \cite{SI}, this will naturally give rise
to a single component ordering parameter that breaks all the mirrors at $T_{CDW}$, and would be consistent with our data. We note that the possibility of a tilted ordering wave-vector has already been discussed as a possible competing ordering in  nesting-driven approaches of the CDW state in RTe$_3$ \cite{yao_theory_2006}. To further confirm this expectation, we computed the CDW AM Raman responses within an effective tight-binding model of $(p_{x^{\prime}},p_{y^{\prime}})$ tellurium orbitals widely used to describe tritellurides \cite{yao_theory_2006,eiter_alternative_2013, alekseev_charge_2024}. The model includes nearest-neighbor hopping both parallel and perpendicular to the $p_{x^{\prime}}$ orbital, with $x^{\prime}=a'$ and 
$y^{\prime}=c'$, as well as next-nearest-neighbor diagonal hopping. The latter induces orbital hybridization and leads to $I_{aa} \ne I_{cc}$ (see SI for additional details \cite{SI}), allowing us to recover a finite $\Sigma_{m'}$, in agreement with the experimental observations. The CDW instability is associated with an ordering wavevector ${\bf Q}_{CDW}$ oriented at an angle $(1+r)\pi/4$ with respect to the $x^{\prime}$ axis (see Fig.\ \ref{fig5}(a)). When $r\ne0$, ${\bf Q}_{CDW}$ deviates from the high-symmetry $c$ direction, resulting in $I^{}_{a'c'} \ne I^{}_{c'a'}$ and the $\Sigma_m \neq 0$. Figures \ref{fig5}(b)-(c) display the calculated differential polarization-resolved Raman responses $I^{}_{x_{\theta}x_{\theta}} - I^{}_{y_{\theta}y_{\theta}}$ and $I^{}_{x_{\theta}y_{\theta}} - I^{}_{y_{\theta}x_{\theta}}$ as a function of the angle $\theta$ between the incident (scattered) light polarization direction $\hat x_{\theta}$ ($\hat y_{\theta}$) and the $a'$ ($c'$) axis. Notice that $I^{}_{x_{\theta}x_{\theta}} - I^{}_{y_{\theta}y_{\theta}}$ reduces to $\Sigma_{m^{\prime}}$ at $\theta=-\pi/4$, while $I^{}_{x_{\theta}y_{\theta}} - I^{}_{y_{\theta}x_{\theta}}$ coincides with $\Sigma_m$ at $\theta=0$.  The theoretical predictions are compared with angle-resolved polarized Raman data on an un-strained sample at 175~K (see SI for the Raman data at 175~K \cite{SI}). The good agreement between experimental data and theoretical predictions when $r\ne0$ further supports the scenario of a tilted ordering vector ${\bf Q}_{CDW}$ as the origin of the simultaneous breaking of the $m$ and $m'$ mirror symmetries. Our proposal should motivate revisiting high resolution X-ray diffraction measurements in search for the proposed tilt and the associated monoclinicity.

In conclusion, our results showcase the power of elasto-Raman spectroscopy to track mirror symmetry breaking transitions under strain. In the context of RTe$_3$ two different types of mirror symmetries are broken at the CDW transition, each of which can be probed independently under varying strain. Our results show that the two resulting order parameters track each other linearly with varying temperature and external strain. The linear relationship indicates that the CDW transition is most likely described by a single-component order parameter, with an ordering wave-vector that is tilted away from a high symmetry direction. The methodology outlined in this work is quite general and should be applicable to other exotic density wave orders.

\section{Acknowledgments}
We acknowledge funding from the Agence National de la Recherche via the grants ANR "SUPER2DTMD" and ANR "Tri-QMat". This work has been supported by Region Île-de-France in the framework of DIM QuanTiP and DIM SIRTEQ. Crystal growth and characterization at Stanford was supported by the Department of Energy, Office of Basic Energy Sciences, under contract DE-AC02-76SF00515. 

\bibliography{bib.bib}
 
\end{document}